\documentclass[10pt,aps,prb,twocolumn,amsmath,notitlepage,floatfix,footinbib,superscriptaddress,showpacs,showkeys]{revtex4-2}
\usepackage{graphicx}
\usepackage{xcolor}
\usepackage{hyperref}

\begin{document}

\title{The Kondo Cloud in a 1D Nanowire}

\author{Joseph Kleinhenz}
\author{Igor Krivenko}
\affiliation{Department of Physics, University of Michigan, Ann Arbor, Michigan 48109, USA}
\author{Guy Cohen}
\affiliation{School of Chemistry, Tel Aviv University, Tel Aviv 69978, Israel}
\author{Emanuel Gull}
\affiliation{Department of Physics, University of Michigan, Ann Arbor, Michigan 48109, USA}

\graphicspath{{figures/}}

\renewcommand*{\Re}{\operatorname{Re}}
\renewcommand*{\Im}{\operatorname{Im}}

\date{\today}

\begin{abstract}
  A recent experiment [Nature {\bf 579}, 210--213 (2020)] probed the extent of the Kondo cloud in 1D by measuring the effect of electrostatic perturbations applied a distance $L$ away from the impurity on $T_K$.
  We study the Kondo cloud in a model proposed to describe this experimental setup, consisting of a single impurity Anderson model coupled to two semi-infinite 1D leads.
  In agreement with the experimental results, we find that $T_K$ is strongly affected by perturbations to the lead within the Kondo cloud.
  We obtain a complementary picture of the Kondo cloud in this system by observing how the Kondo state manifests itself in the local density of states of the leads, which may be observed experimentally via scanning tunneling microscopy.
  Our results support the existing experimental data and provide detailed predictions for future experiments seeking to characterize the Kondo cloud in this system.
\end{abstract}

\maketitle

\section{Introduction}
The Kondo effect, first observed in magnetic impurities embedded in metals, is characterized by the screening of an impurity spin by a cloud of conduction electrons in such a way that a singlet state is formed \cite{hewsonKondoProblemHeavy2009}.
This occurs at temperatures below the Kondo temperature $T_K$.
The equilibrium physics of the Kondo effect are by now well understood, but the spatial structure of the screening cloud itself has remained elusive \cite{affleckKondoScreeningCloud2010}.

The Kondo cloud can be defined in a number of ways.
Theoretically, perhaps the most natural definition is in terms of the correlation function between the impurity spin and a conduction electron spin.
This can be readily computed using a variety of techniques including numerical renormalization group (NRG) \cite{bordaKondoCloudSpinspin2009,bordaKondoScreeningCloud2007} and matrix product state \cite{holznerKondoScreeningCloud2009} methods.
However, this correlation function has so far proven difficult to observe experimentally.
Attempts to access it via the magnetic susceptibility in NMR have so far been unsuccessful \cite{boyceConductionElectronSpinDensity1974}.
A number of alternative theoretical \cite{busserNumericalAnalysisSpatial2010, ribeiroNumericalStudyKondo2019, affleckFriedelOscillationsKondo2008, ujsaghyTheoryFanoResonance2000, antipovIdentifyingKondoOrbitals2013} and experimental \cite{liKondoScatteringObserved1998, madhavanTunnelingSingleMagnetic1998, pruserLongrangeKondoSignature2011} efforts were made to characterize the Kondo cloud via its effect on the conduction electron density of states.
The latter can be measured spectroscopically by scanning tunneling microscopy (STM) experiments.
These studies have revealed various signatures of the Kondo cloud, but did not reveal a spatial structure fully consistent with theoretical predictions.
We also note in passing that spin--spin correlations do not necessarily imply a singlet state, and that more stringent (if also complicated) measurements have been proposed \cite{erpenbeckResolvingNonequilibriumKondo2020}.

In \cite{parkHowDirectlyMeasure2013}, the authors proposed a different route: to study the Kondo cloud by examining the effect of perturbations to the conduction electrons on the impurity.
Perturbations of electrons inside the Kondo cloud should have a large effect on the Kondo resonance, whereas perturbations of electrons outside the Kondo cloud should have little effect.
By varying the location at which the perturbation is applied, it is then possible to map out the Kondo cloud in space.
This proposal was recently realized experimentally by applying electrostatic perturbations to a 1D channel coupled to a quantum dot (QD).
By observing the effect of perturbations on electronic transport, the extent of the Kondo cloud was successfully measured and found to agree with theory \cite{borzenetsObservationKondoScreening2020}.

In this paper we study a model proposed to describe this experimental setup.
We characterize the Kondo cloud in this model in two complementary ways.
First, we compute the effect of lead perturbations applied a variable distance $L$ away from the impurity on the Kondo resonance width, which is related to the Kondo temperature.
Our results confirm the general scenario found in \cite{borzenetsObservationKondoScreening2020}, in which lead perturbations cause Kondo temperature fluctuations which decay with $L$.
At large interaction strengths, we find a non-monotonic dependence of the Kondo resonance width on $L$ which may be observable in future experiments.
Additionally, we compute the lead local density of states (LDOS), observe features associated with the emergence of the Kondo state, and show that these features are suppressed by applying a nonequilibrium bias voltage.
These results provide a detailed picture of the Kondo cloud in this system that can guide future STM experiments that will shed light on Kondo physics in a wide range of equilibrium and nonequilibrium scenarios.

\section{Model}
\begin{figure}
  \includegraphics[width=\columnwidth]{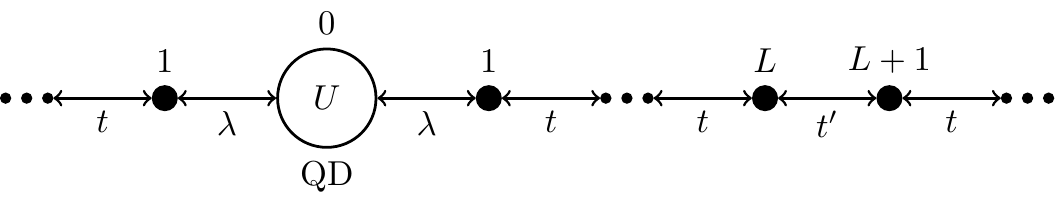}
  \caption{\label{fig:setup}
  Schematic illustration of the model.
  A quantum dot (QD) is coupled to two noninteracting leads, where each lead is a semi-infinite, one-dimensional tight-binding chain with nearest neighbor hopping $t$.
  An on-site local Coulomb interaction $U$ is present on the impurity site.
  The leads are connected to the QD by a hopping $\lambda$.
  On the right lead, the hopping between sites $L$ and $L+1$ is assumed to be modifiable to take the value $t'$.
  }
\end{figure}
We study the model proposed in \cite{borzenetsObservationKondoScreening2020} to describe an experimental setup for observing the Kondo cloud.
The model consists of a single orbital QD coupled to two noninteracting, 1D leads (see Fig.~\ref{fig:setup}).
The Hamiltonian for this model is
\begin{align}
  H &= H_\mathrm{QD} + H_l + H_r + H_{T},
\end{align}
where $H_\mathrm{QD}$ is the quantum dot Hamiltonian, $H_l$ ($H_r$) is the Hamiltonian of the left (right) lead, and $H_T$ is the tunneling Hamiltonian that describes hopping between the QD and the leads.

The QD Hamiltonian is
\begin{align}
  H_\mathrm{QD} &= \sum_\sigma \epsilon_d n_\sigma + U n_\uparrow n_\downarrow,
\end{align}
where $d^\dagger_\sigma$ ($d_\sigma$) creates (annihilates) electrons localized on the QD with spin $\sigma$; $n_\sigma = d^\dagger_\sigma d_\sigma$ is the QD number operator; $\epsilon_d$ is the single-particle energy; and $U$ is the Coulomb interaction between electrons on the QD.

The left lead is modeled as a uniform 1D tight-binding chain with Hamiltonian
\begin{align}
  H_l = -\sum_\sigma \sum_{i = 1}^{\infty} t c^\dagger_{l,i,\sigma} c_{l,\left(i+1\right),\sigma} + \mathrm{h.c.}
\end{align}
Here, $c^\dagger_{l,i,\sigma}$ ($c_{l,i,\sigma}$) creates (annihilates) electrons on site $i$ of lead $l$ with spin $\sigma$, and $t$ is the nearest-neighbor hopping amplitude.

The right lead is identical to the left lead, except in that the hopping amplitude between sites $L$ and $L+1$ is set to $t'$ rather than $t$.
When $t'<t$, transport between these two sites is reduced, partially pinching off the lead at site $L$.
This can be written as follows:
\begin{align}
  \begin{split}
    H_r &= -\sum_\sigma \biggl[ \sum_{i\ne L} t c^\dagger_{r,i,\sigma} c_{r, \left(i+1\right),\sigma} \\
        &\qquad+ t' c^\dagger_{r,L,\sigma} c_{r,\left(L+1\right),\sigma} + \mathrm{h.c.} \biggr],
  \end{split}
\end{align}
where $c^\dagger_{r,i,\sigma}$ ($c_{r,i,\sigma}$) creates (annihilates) electrons on site $i$ of lead $r$ with spin $\sigma$.
We specify the strength of the modification in terms of the dimensionless parameter $\alpha = 1 - \left(t'/t\right)^2$, which continuously interpolates between a case with no modification ($\alpha = 0$) and completely pinching off the lead at $L$ ($\alpha = 1$).

Finally, the coupling of the QD to the leads is described by the tunneling Hamiltonian
\begin{align}
  H_T &= \sum_\sigma \sum_{w=l,r} \lambda_w \left(c^\dagger_{w,1,\sigma} d_\sigma + \mathrm{h.c.}\right)
\end{align}
where $\lambda_w$ is the hopping amplitude between the QD and lead $w$.

The leads are asssumed to be at half-filling.
We take the spacing between the sites to be $a$, which we use as our unit of distance.
Given this, the Fermi wavevector is $k_f = \pi/2a$ and the bulk Fermi velocity $v_f = 2 t a / \hbar$.
When the leads are decoupled from the QD, the electronic dynamics within them are governed by a noninteracting Hamiltonian and their Green's functions can be formally computed from
\begin{align}
  g_w(\omega) = \left(\omega I - H_w\right)^{-1}.
\end{align}
Here, $I$ is the identity matrix and $H_w$ is the single-particle Hamiltonian matrix within lead $w$; we use the same notation as for the corresponding many-body Hamiltonian to minimize terminology, since the intention is clear from context.
The calculation of the lead Green's functions is described in detail in appendix \ref{app:derivation}.

\begin{figure}
  \includegraphics[]{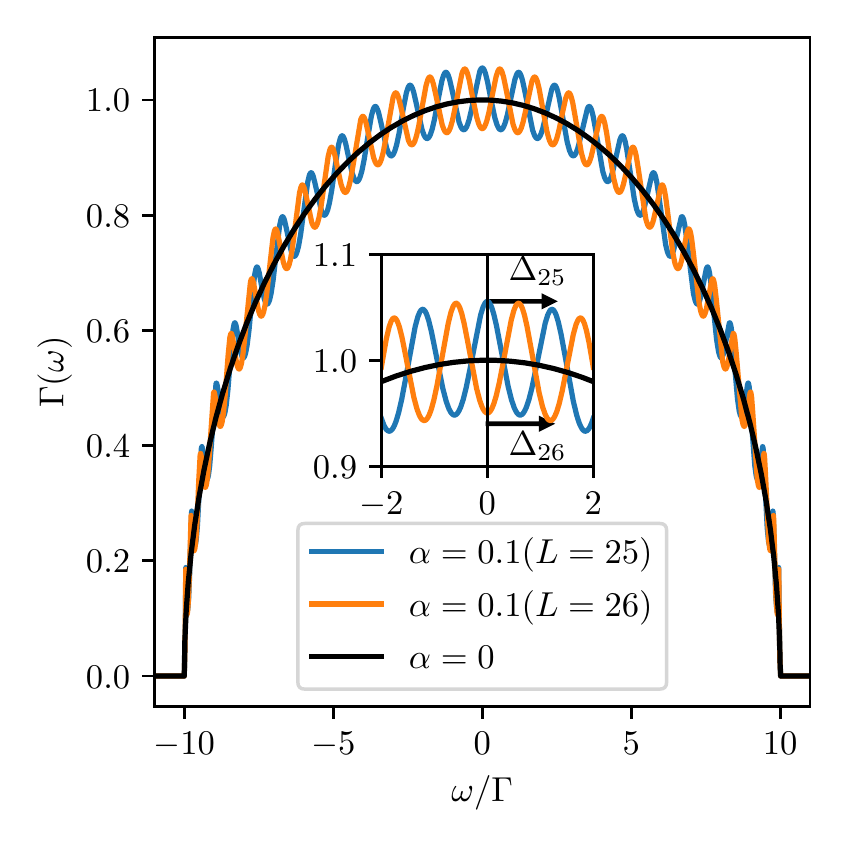}
  \caption{\label{fig:hyb}
  Coupling density $\Gamma(\omega)$ for $\alpha = 0$ (black) and $\alpha \ne 0$ for two different cavity lengths $L$ (blue/orange).
  For $\alpha \ne 0$ the resonance width is given by $\Delta_L = \pi v_f / L$.
  The inset focuses on the region near the Fermi energy, illustrating the effect of nonzero $\alpha$.
  }
\end{figure}
Because the leads are noninteracting, they can be integrated out exactly to obtain a hybridization function (or embedding self-energy) that describes their effect on the QD.
This hybridization function parameterizes the lead band structure and the tunneling Hamiltonian together.
It can be fully defined in terms of the coupling density
\begin{align} \label{eq:hyb}
  \Gamma(\omega) = - \Im \sum_{w=l,r} \lambda_w^2 g_{w,11}(\omega),
\end{align}
where $g_{w,11}$ is the local Green's function of the site on lead $w$ adjacent to the quantum dot \cite{cuevasMolecularElectronicsIntroduction2017}.
A procedure for computing $g_{w,11}$ is given in appendix \ref{app:derivation}.
In the uniform ($\alpha = 0$) case, the coupling density is semi-circular with half-bandwidth $2t$:
\begin{align}
  \Gamma(\omega)
  = \sum_{w=l,r}\frac{\lambda_w^2}{2 t^2}
  \begin{cases}
    \sqrt{4t^2 - \omega^2} & \left|\omega\right| \le 2t, \\
    0 & \left|\omega\right| > 2t.
  \end{cases}
\end{align}
In the pinched case ($\alpha > 0$), the reduced hopping between sites $L$ and $L+1$ creates a Fabry--P\'erot cavity in the right lead with resonance width $\Delta = \pi v_f / L$.
Changing $L$ switches the cavity between on- and off-resonance ($e^{2 i k_f L} = \pm 1$) and flips $\Gamma(0)$ between being a minimum and being a maximum, respectively.
Fig.~\ref{fig:hyb} shows the coupling density for both the uniform and pinched cases.
The overall hybridization strength is parameterized by the $\alpha = 0$ level broadening, $\Gamma = \Gamma(\omega = 0) = \sum_w \lambda_w^2/t$.
We use $\Gamma$ as our unit of energy, and set $\hbar \equiv 1$.

An important property of this model is its Kondo temperature $T_K$.
At temperatures below $T_K$, the QD spin is screened by the lead electrons to form a singlet state.
The bare cloud length $\xi_0 = v_f / T_K$ is the theoretically expected spatial extent of the cloud of lead electrons that make up this singlet state \cite{affleckKondoScreeningCloud2010}.

Ref.~\onlinecite{borzenetsObservationKondoScreening2020} provides detailed experimental parameter estimates for this model, which we use as a guide in choosing our parameter values.
The level broadening $\Gamma$ is estimated to be approximately $0.1\thinspace\textrm{meV}$, which implies a unit of temperature given by $\Gamma / k_B \approx 1.16\thinspace\textrm{K}$.
Following Ref.~\cite{borzenetsObservationKondoScreening2020}, we choose $\alpha = 1 - \left(t'/t\right)^2 = 0.1$ for all cases with modified hopping.
In the experiment, a coupling asymmetry given by $\lambda_r \approx 4 \lambda_l$ is suggested, although this is tunable.
For simplicity, we choose $\lambda_l = \lambda_r = \lambda$ so that the leads are symmetrically coupled to the QD.
Note that our choice of energy unit, $\Gamma = 1$, implies $\lambda = \sqrt{t/2}$.

The experimental estimates suggest $U \approx 6 \Gamma$.
In our calculations we consider slightly larger interactions ($U = 7\Gamma,\, 8\Gamma,\, 9\Gamma$) in order to be in a regime where the approximation we will use is more accurate.
Also as in the experiment, we consider only the symmetric situation where $\epsilon_d = -U/2$ so that each state in the isolated QD is doubly degenerate in energy.
The model is also spin-symmetric, and spin indices on Green's functions and observables will therefore be omitted.
We choose the lead half-bandwidth $D = 2t = 10\Gamma$.
This follows the experimental parameter estimates in making the lead half-bandwidth the largest scale in the problem, while reducing it from the experimental value $(D \approx 60 \Gamma)$ for computational convenience.
The experimental Fermi velocity is estimated to be $v_f \approx 2.5 \times 10^5\thinspace\textrm{m/s}$ which implies $a = v_f / (2t) \approx 150\thinspace\textrm{nm}$.
Note that within the mesoscopic realization, $a$ should be thought of as a phenomenological parameter rather than the spacing between physical atoms in the system.

The first quantity that will be of interest to us is the QD density of states (DOS)
\begin{align}
  \rho_\mathrm{imp}(\omega) = -\frac{1}{\pi} \Im G_\mathrm{imp}(\omega),
\end{align}
where $G_\mathrm{imp}(\omega)$ is the frequency-dependent retarded QD Green's function.
The second quantity of interest is the site-dependent local density of states (LDOS) of the leads
\begin{align}
  \rho_w(\omega, i) = -\frac{1}{\pi} \Im G_{w, ii}(\omega),
\end{align}
where $G_{w, ii}(\omega)$ is the frequency-dependent retarded local Green's function of site $i$ of lead $w$ (governed by the full Hamiltonian).
This Green's function can be obtained from the QD Green's function using the relation
\begin{align} \label{eq:leadG}
  G_{w,ii}(\omega) = g_{w,ii}(\omega) + \lambda^2 g_{w,i1}(\omega) G_\mathrm{imp}(\omega) g_{w,1i}(\omega),
\end{align}
where $g_{w}$ is the noninteracting Green's function of lead $w$.
A derivation of this result is given in appendix \ref{app:derivation}.

\section{Methods}
Most studies of Kondo physics---including previous work on the present model---rely on what arguably remains the most efficient and reliable methodology for accessing low energy physics in the Kondo regime: the numerical renormalization group (NRG) \cite{wilsonRenormalizationGroupCritical1975, bullaNumericalRenormalizationGroup2008,parkHowDirectlyMeasure2013, borzenetsObservationKondoScreening2020}.
Here, we will focus on frequency correlation functions, higher temperatures above the scaling regime and nonequilibrium effects.
Much of the focus is therefore on higher-energy physics.
To compute the QD Green's function $G_\mathrm{imp}(\omega)$, we employ diagrammatic expansions formulated on the three-branch (forward, backward and imaginary) Keldysh contour.
Using these methods the QD's time-dependent retarded Green's function can be written as follows at the steady state:
\begin{align}
  G_\mathrm{imp}(t - t') = -i \left\langle \left\{d_\sigma(t), d_\sigma^\dagger(t')\right\} \right\rangle.
\end{align}
In practice, within our methodology the dynamics can be computed only up to some finite maximum time, such that we always have $t,t' \le t_\mathrm{max}$.
The frequency dependent Green's function is then obtained by a Fourier transform
\begin{align}
  G_\mathrm{imp}(\omega) = \int_0^{t_\mathrm{max}} dt\, e^{i \omega t} G_\mathrm{imp}(t)
\end{align}
and all further analysis is performed in the frequency domain; $t_\mathrm{max}$ therefore sets a limit on our possible frequency resolution.
We also note that in the case where a nonzero voltage is applied, the system begins in a state that differs from the final steady state, and exact results (even at large frequencies) are only obtained at the limit of large $t_\mathrm{max}$.

Inchworm quantum Monte Carlo (QMC) is a numerically exact method for evaluating population dynamics \cite{cohenTamingDynamicalSign2015} and Green's functions \cite{antipovCurrentsGreenFunctions2017}.
It is based on a Keldysh expansion in the tunneling Hamiltonian $H_T$, also known as the hybridization expansion \cite{wernerContinuousTimeSolverQuantum2006,muhlbacherRealTimePathIntegral2008,wernerDiagrammaticMonteCarlo2009,schiroRealtimeDiagrammaticMonte2009,schiroRealtimeDynamicsQuantum2010}.
Inchworm QMC methods have been applied to a variety of applications involving both quantum impurity models \cite{boagInclusionexclusionPrincipleManybody2018,ridleyNumericallyExactFull2018,krivenkoDynamicsKondoVoltage2019,ridleyNumericallyExactFull2019,ridleyLeadGeometryTransport2019,eidelsteinMultiorbitalQuantumImpurity2020} and strongly correlated materials \cite{dongQuantumMonteCarlo2017,kleinhenzDynamicControlNonequilibrium2020}.
Since the method is computationally demanding, it is not yet feasible to reach long enough $t_\mathrm{max}$ to resolve fine spectral features without broadening.
Because of this, we base the bulk of our study on the computationally less expensive one crossing approximation (OCA).
This corresponds to a second-order truncation of the dressed hybridization series used in the inchworm expansion, and can be implemented semi-analytically \cite{pruschkeAndersonModelFinite1989, ecksteinNonequilibriumDynamicalMeanfield2010}.
The OCA becomes increasingly accurate at large $U$.
With the appropriate vertex corrections, the OCA accounts well even for some aspects within the low temperature scaling regime \cite{andersPerturbationalApproachAnderson1994,andersNCANewResults1995,greweConservingApproximationsDirect2008}, though we do not employ such corrections or explore this regime here.
We do, however, validate our OCA results against numerically exact inchworm QMC results in the parameter regime where this is feasible (see appendix \ref{app:inch}).

\section{Results}

\begin{figure}
  \includegraphics[]{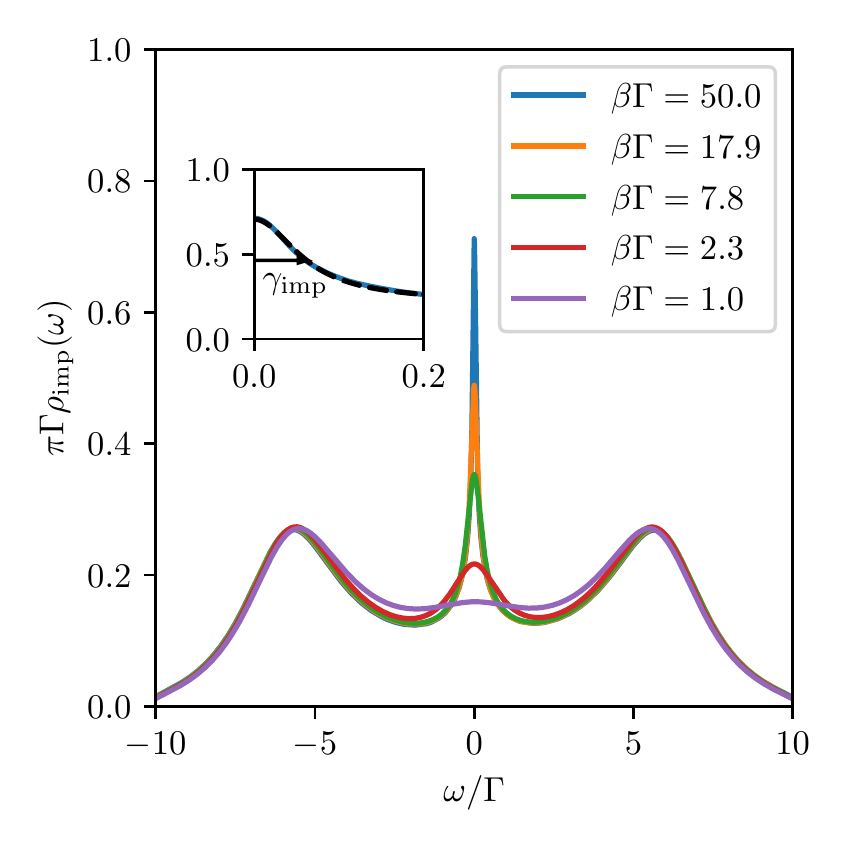}
  \caption{\label{fig:spectrum_imp}
  Impurity DOS at interaction strength $U = 9\Gamma$ as the temperature is lowered below the Kondo temperature $T_K$.
  The inset zooms in on the peak at the Fermi energy for the $\beta = 50\Gamma^{-1}$ case.
  The dashed black line shows a Lorentzian fit used to estimate the width $\gamma_\mathrm{imp} \approx 0.062\Gamma$, which provides us with an estimate for the Kondo temperature $T_K$.
  In these units, the zero temperature unitary limit corresponds to $\pi \Gamma \rho_{\mathrm{imp}}(0) = 1$.
  }
\end{figure}
Fig.~\ref{fig:spectrum_imp} shows the impurity DOS at $U = 9\Gamma$ for a sequence of inverse temperatures $(\beta = \frac{1}{k_B T})$ between $\beta \Gamma = 1$ and $\beta \Gamma = 50$.
These results are computed with $\alpha = 0$, in the absence of any cavity.
As the temperature is decreased below the Kondo temperature $T_K$, the impurity spectrum builds up a sharp Kondo peak at the Fermi energy with width $\gamma_\mathrm{imp}$.
We obtain the width of the peak by fitting a Lorentzian (with an additional offset term) to the spectrum around the Fermi energy.
The fit function is given by
\begin{align}
  f(\omega) = A \frac{1}{\omega^2 + \gamma_\mathrm{imp}^2} + B.
\end{align}
Here, $A$ and $B$ are fit parameters and $\gamma_\mathrm{imp}$ is the estimated width of the Kondo peak.
The inset of Fig.~\ref{fig:spectrum_imp} shows the impurity spectrum (blue line) around the Fermi energy together with the fit (dashed black line).

As $T \to 0$, $\gamma_\mathrm{imp}(T)$ converges to the Kondo temperature and $\rho_\mathrm{imp}(0)$ converges to $\frac{1}{\pi\Gamma}$ (in the wide band limit) \cite{nagaokaTemperatureDependenceSingle2002, hewsonKondoProblemHeavy2009, darocaRelationWidthZerobias2018}.
At $\beta \Gamma = 50$, $\gamma_\mathrm{imp}$ is not fully converged to the zero temperature value and therefore overestimates $T_K$.
Nevertheless it still provides a useful estimator that tracks changes in the Kondo temperature.
For $U = \left\{7\Gamma,\, 8\Gamma,\, 9\Gamma\right\}$ we estimate inverse widths of $1/\gamma_{\mathrm{imp}} = \left\{13.3\Gamma^{-1},\, 14.8\Gamma^{-1},\, 16.1\Gamma^{-1}\right\}$, respectively.
Note that this procedure is different from the method for estimating the Kondo temperature used in Ref.~\onlinecite{borzenetsObservationKondoScreening2020}, which defines $T_K$ as the temperature at which the conductance reaches half of its zero temperature value \cite{goldhaber-gordonKondoRegimeMixedValence1998}.
Since $T_K$ defines a crossover scale rather than a sharp transition, its exact value is ambiguous.
For example, in \cite{darocaRelationWidthZerobias2018}, the authors find that within the noncrossing approximation (NCA) the $T_K$ estimated from the impurity spectrum is approximately half the $T_K$ estimated from the conductance.

\begin{figure}
  \includegraphics[]{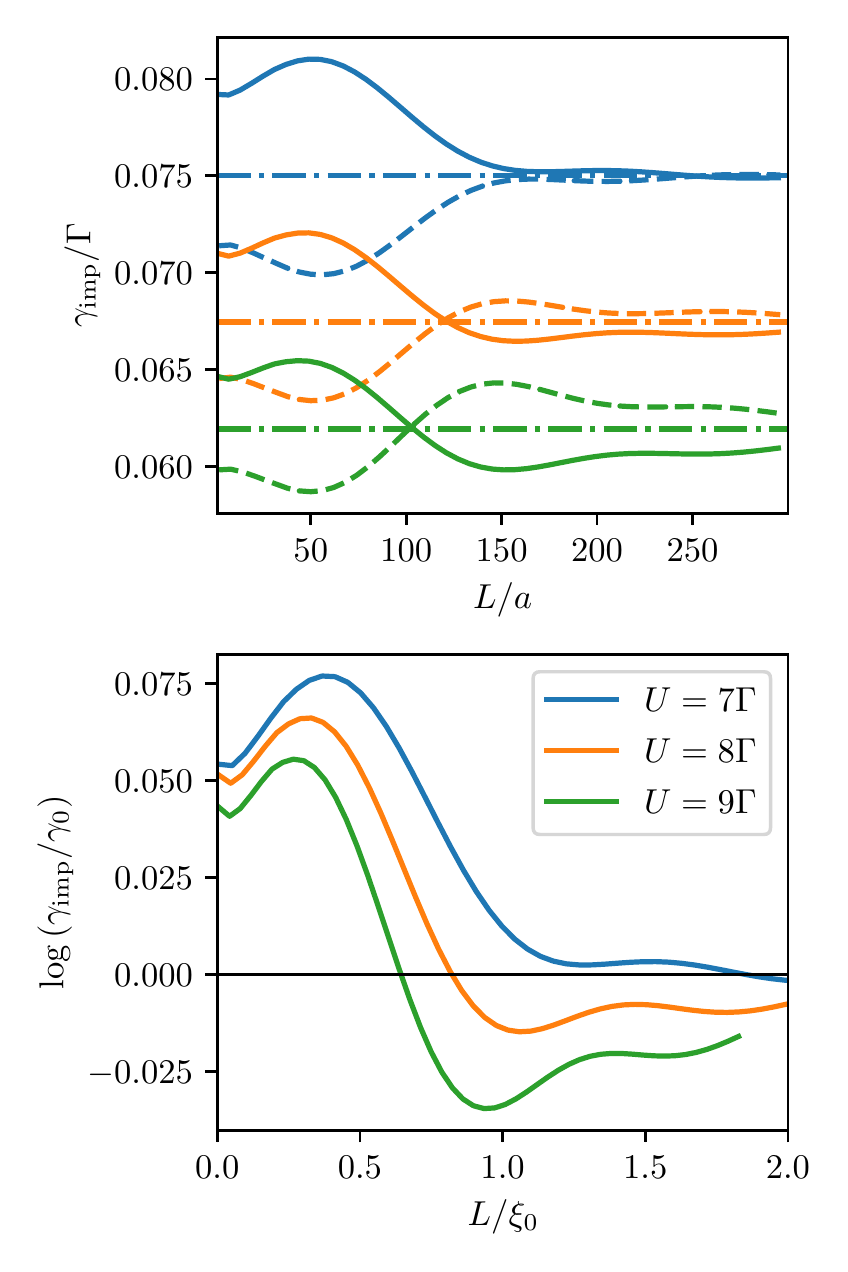}
  \caption{\label{fig:pinch}
  Top: Kondo peak width $\gamma_\mathrm{imp}$ as a function of the size of the Fabry-P\'erot cavity length $L$ for $\alpha = 0.1$ and $U = \left\{7\Gamma,\,8\Gamma,\,9\Gamma\right\}$.
  Solid (dashed) lines show data for $L$ odd (even).
  Horizontal lines show data in the absence of the cavity ($\alpha = 0$).
  Bottom: Logarithm of peak width $\gamma_\mathrm{imp}$ normalized by peak width in the absence of the cavity $\gamma_0$.
  }
\end{figure}
We now consider the effect of lead perturbations ($\alpha = 0.1$) a distance $L$ away from the impurity.
We study how the perturbations change the width $\gamma_{\mathrm{imp}}$ which we take as a proxy for changes in the Kondo temperature $T_K$.
In the experimental setup $L$ is varied on the order of the Fermi wavelength around three fixed distances.
The experimental parameter estimates of the resonance width give $\Delta = \pi v_f / L \approx \left\{3\Gamma,\, 1.2\Gamma,\, 0.75\Gamma\right\}$, corresponding to $L = \pi v_f / \Delta \approx \left\{10a,\, 26a,\, 42a\right\}$.
This implies a maximum experimental $L$ on the order of twice the bare cloud length estimated from the theoretical formula $\xi_0 = v_f / T_K$.

The top panel of Fig.~\ref{fig:pinch} shows our results for the Kondo peak width $\gamma_\mathrm{imp}$ as a function of $L$.
For scale, $v_f / \gamma_{\mathrm{imp}} = \left\{133a,\,148a,\,161a\right\}$ for $U = \left\{7\Gamma,\,8\Gamma,\,9\Gamma\right\}$ respectively.
Note $v_f / \gamma_{\mathrm{imp}}$ underestimates the bare cloud length $\xi_0 = v_f / T_K$, since at $\beta \Gamma = 50$ we expect that $\gamma_{\mathrm{imp}}$ be an overestimate for $T_K$ due to thermal broadening.
The width shows a pronounced even--odd effect which comes from switching the cavity between on- and off-resonance states ($e^{2 i k_f L} = \pm 1$).
The bottom panel of Fig.~\ref{fig:pinch} shows $\gamma_\mathrm{imp}$ for odd sites only, normalized by the peak width in the absence of the cavity.
These results agree with the results of Refs.~\cite{borzenetsObservationKondoScreening2020, parkHowDirectlyMeasure2013} in predicting Kondo temperature oscillations which decay with $L$.
The magnitude of the oscillations is somewhat smaller than in the experiment.
Two factors account for this difference.
The first is thermal broadening of the Kondo peak.
The second is our choice of using a symmetric coupling to the leads, instead of having a stronger coupling to the perturbed lead.

These results provide a detailed picture of the effect of perturbing the lead.
Notably, the amplitude of the width oscillations is a non-monotonic function of the distance $L$.
For small $L$ $(\lesssim 50a)$ the oscillation amplitude slowly increases with $L$.
For larger $L$ $(\lesssim 100a)$ the oscillation amplitude decays linearly.
Interestingly, for $L \gtrsim 100a$ the behavior depends on the value of $U$.
For $U = 7\Gamma$ (the case closest to the experimental value) the oscillation amplitude simply decays and remains very small as $L$ is increased.
However, for larger $U$, the lines for even and odd $L$ cross each other and the even--odd effect flips in direction at large $L$.
This crossover should be observable in future experiments.
The oscillation amplitude then flattens off around $L \simeq 150a$ and begins to exhibit a slow decay.

A non-monotonic dependence of $T_K$ on $L$ was predicted in Ref.~\cite{parkHowDirectlyMeasure2013} using Anderson's poor man's scaling technique \cite{andersonPoorManDerivation1970}, but not in NRG.
The authors attributed the monotonic dependence seen in their NRG results to a failure of the logarithmic discretization to fully resolve the energy scale $\Delta$ introduced by the lead perturbation.
Our results provide further evidence that the expected dependence is non-monotonic, and that it may not be fully captured by the logarithmic discretization employed in traditional NRG methods.
It would be interesting to revisit this problem with newer NRG methods that allow a more flexible band discretization \cite{weichselbaumVariationalMatrixproductstateApproach2009}.

Due to the complicated, non-monotonic behavior, it is difficult to extract a numerical value for the Kondo cloud length from the data shown here.
The results clearly reveal that the lead perturbation has a pronounced effect on the Kondo state, which shows an overall decay with $L$ and reveals something about the nature of the Kondo cloud.
However, the interpretation of this behavior in order to extract a length scale characterizing the cloud remains difficult, especially at large $U$.
Because of this, it is interesting to consider other experimental modalities that could be applied to the same system in order to obtain a complementary view of the Kondo cloud.

\begin{figure}
  \includegraphics[]{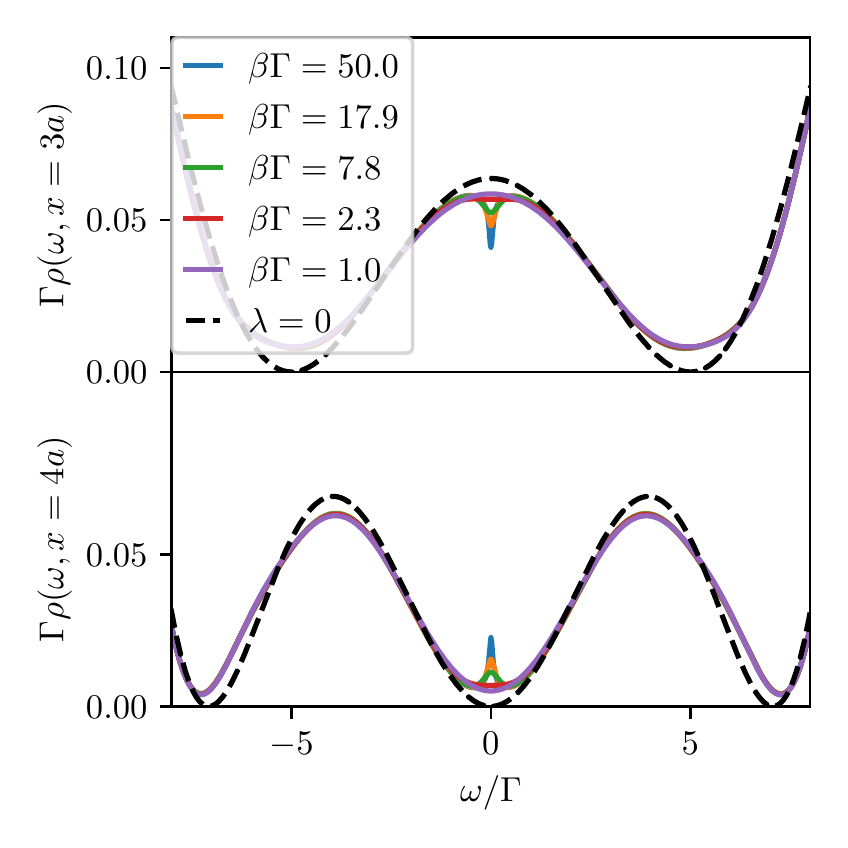}
  \caption{\label{fig:spectrum_lead}
  Lead LDOS at interaction strength $U = 9\Gamma$ at sites 3 (top) and 4 (bottom), for a series of temperatures spanning $T_K$.
  The dashed black line shows the LDOS for a noninteracting lead decoupled from the impurity.
  }
\end{figure}
A promising alternative approach is to consider the local density of states (LDOS) in the lead, which is experimentally accessible in STM experiments.
Since a cavity or perturbation in the lead is no longer required when the LDOS is available, we now consider $\alpha = 0$.
Fig.~\ref{fig:spectrum_lead} shows the lead LDOS at two sites on the right lead at several temperatures.
When the temperature dips below the Kondo scale $T_K$, a Kondo peak emerges at the Fermi energy in the QD's DOS, and a corresponding feature emerges at the Fermi energy in the lead LDOS.
For odd sites, this feature is seen as a dip around the Fermi energy relative to the high temperature spectrum, whereas for even sites the feature is seen as a peak.

\begin{figure} 
  \includegraphics[]{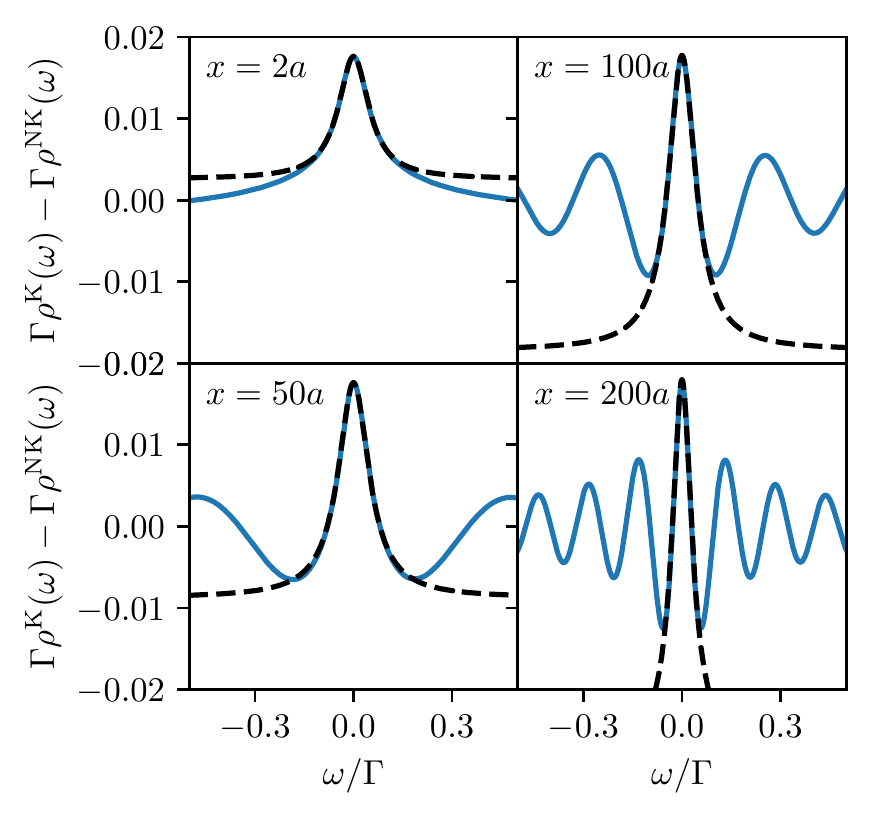}
  \caption{\label{fig:spectrum_lead_diff}
  Difference between the lead LDOS above and below $T_K$ on four different sites.
  The dashed black line shows a Lorentzian fit of the central peak, used to estimate $\gamma(x)$.
  }
\end{figure}
In order to observe how the Kondo cloud manifests in the lead LDOS we compare the non-Kondo LDOS $\rho^{NK}(\omega)$, which we observe at $T = 1\Gamma \gg T_K$, to the Kondo LDOS $\rho^{K}(\omega)$ observed at $T = 0.02 \Gamma \lesssim T_K$.
Fig.~\ref{fig:spectrum_lead_diff} shows the difference $\rho^K(\omega) - \rho^{NK}(\omega)$ for four different sites.
The difference is largest at the Fermi energy, decays rapidly with $\omega$ and becomes oscillatory with increasing distance from the QD.
This approach to characterization of the Kondo cloud is related to what has previously been explored in Refs.~\onlinecite{busserNumericalAnalysisSpatial2010, ribeiroNumericalStudyKondo2019}.

Ref.~\onlinecite{busserNumericalAnalysisSpatial2010} proposed measuring the extent of the Kondo cloud by examining the function
\begin{align}
  F(n) = \int d\omega \left[\rho^{K}_n(\omega) - \rho^{NK}_n(\omega)\right] L_{\gamma}(\omega),
\end{align}
where $L_{\gamma}(\omega)$ is a Lorentzian with width given by the width of the Kondo peak on the impurity.
This proposal successfully produces an observable that appears to measure the extent of the Kondo cloud.
However, it has the unfortunate feature of directly inserting the Kondo temperature (via the impurity DOS width $\gamma$) into the measurement of the Kondo cloud.
Ideally, one would like to have a measurement of the Kondo cloud which is as independent as possible from other measurements in order to be able to check its scaling properties.

In Ref.~\onlinecite{ribeiroNumericalStudyKondo2019}, it is instead proposed that the extent of the Kondo cloud be measured by examining the function
\begin{align}
  L(n) = \int d\omega \left|\rho^{K}_n(\omega) - \rho^{NK}_n(\omega)\right|,
\end{align}
which integrates the absolute difference between the Kondo and the non-Kondo spectra over the entire bandwidth.
This procedure avoids inserting the Kondo temperature into the measurement.
However, it also introduces experimental difficulties.
Since the magnitude of the difference becomes very small and highly oscillatory away from the Fermi energy, high precision measurements over the entire energy window are required.

As an alternative to these methods, we propose to measure the Kondo cloud in the lead LDOS by looking at the width of the peak/dip at the Fermi energy.
This avoids both unnecessarily introducing $T_K$ into the measurement, and the need for extremely high precision measurements over the entire bandwidth.
The black dashed lines in Fig.~\ref{fig:spectrum_lead_diff} show fits of a Lorentzian to the central peak, which we use to extract the width $\gamma(x)$.
From the figure we see that a Lorentzian provides a good fit of the central peak and that the peak narrows with distance from the QD.

\begin{figure}
  \includegraphics[]{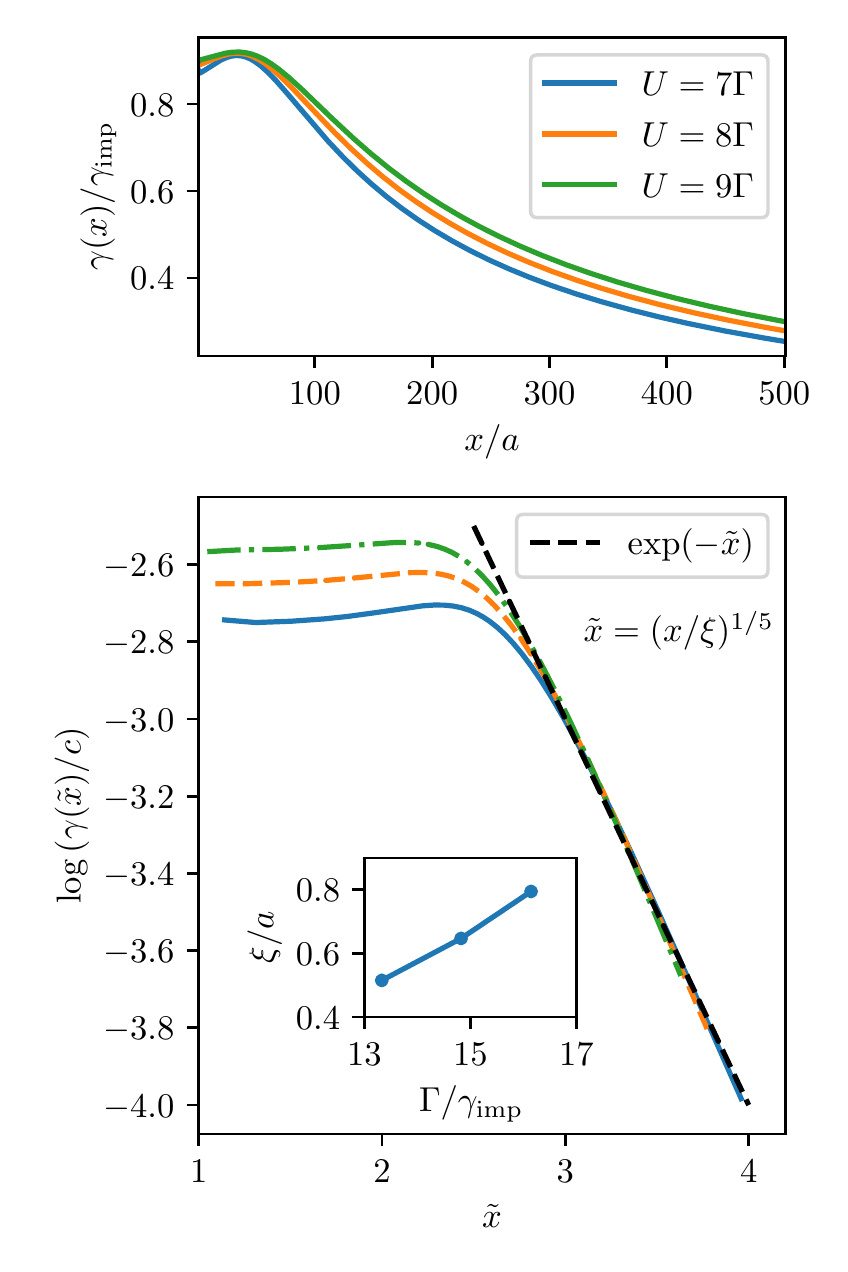}
  \caption{\label{fig:width}
  Top: Width of lead LDOS normalized by the impurity width for $U = \left\{7\Gamma,\, 8\Gamma,\, 9\Gamma\right\}$.
  Bottom: Logarithm of LDOS width as a function of the rescaled distance.
  Inset: Extracted length scale $\xi$, which varies inversely with $\gamma_\mathrm{imp} \approx T_K$.
  }
\end{figure}
The top panel of Fig.~\ref{fig:width} shows the dependence of the LDOS width $\gamma(x)$ on distance from the QD for $U = \left\{7\Gamma,\,8\Gamma,\,9\Gamma\right\}$.
Note this data contains both even and odd sites, demonstrating that there is no even--odd effect in the width.
For small $x$ the data shows $\gamma(x)$ slightly increasing.
As $x$ increases, $\gamma(x)$ begins a smooth monotonic decay.
It is worth noting that this behavior appears much simpler than the behavior of $T_K$ as a function of $L$ in the presence of a cavity.

The bottom panel of Fig.~\ref{fig:width} shows that $\gamma(x)$ appears to demonstrate universal scaling behavior.
To extract the length scale $\xi$ we fit a function $f(x) = c \exp{\left[-\left(x/\xi\right)^{1/5}\right]}$ to $\gamma(x)$ over the range $100a \le x \le 500a$ for each $U$.
The particular form of this fit function was empirically determined.
In particular, the exponent was initially a free parameter but was found to take values $\approx 1/5$ and was then fixed.
The plot shows the curve collapse generated by plotting $\log\left(\gamma(\tilde{x})/c\right)$ against $\tilde{x} = \left(x/\xi\right)^{1/5}$.
The robust linear behavior for $\tilde{x} \gtrsim 3$ shows that this correctly describes the scaling behavior far from the QD.

In the inset of Fig.~\ref{fig:width} we plot the extracted length scale $\xi$ against the inverse Kondo peak width.
As expected, this shows linear behavior consistent with the theoretical relationship $\xi_0 = v_f/T_K$.
Note that the values of $\xi$ are small relative to the expected length scales of ${\sim} 100a$.
However, this small value should not be taken to imply that there is no Kondo cloud.
The small values of the length scale $\xi$ describing the asymptotic decay are due to the small exponent of our fit function.
The top panel of Fig.~\ref{fig:width} shows that the width remains a substantial fraction of the impurity peak width over length scales of ${\sim} 100a$.

This analysis offers a view of the Kondo cloud complementary to that provided by the lead perturbation method.
In particular, the expected functional dependence is simpler and the process of extracting a length scale more straightforward.
However, this method does require high precision STM measurements.
For our parameters, the peak widths are on the order of $0.05\Gamma \approx 5\thinspace \mu \textrm{eV}$.

\begin{figure}
  \includegraphics[]{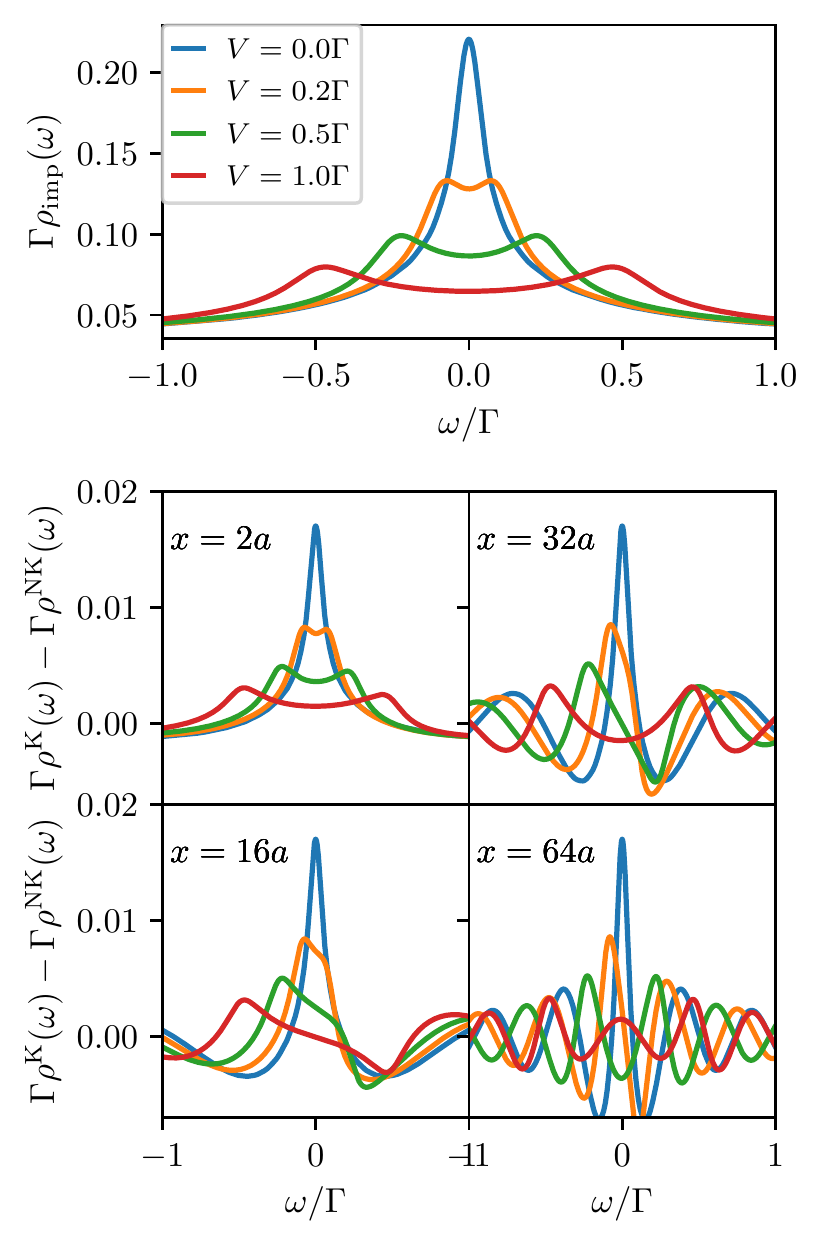}
  \caption{\label{fig:voltage}
  Top: Impurity DOS for several bias voltages $V = \{0.0\Gamma,\,0.2\Gamma,\,0.5\Gamma,\,1.0\Gamma\}$.
  Bottom: Difference between the lead LDOS at $\beta \Gamma = 1$ and $\beta \Gamma = 50$ for each bias voltage, at four different lead sites.
}
\end{figure}

Finally, we consider the effect of applying a bias voltage $V$ between the two leads on the Kondo cloud.
At time $t=0$ we instantaneously shift the chemical potential of the left (right) lead by $+V/2$, $(-V/2)$ and evolve the system to a steady state.
Fig.~\ref{fig:voltage} shows the steady state DOS at the QD (top panel) and the lead LDOS (bottom panel) at several voltages.
On the impurity, applying a voltage suppresses the Kondo peak and splits it, creating two partially suppressed peaks at $\omega \approx \pm V/2$ \cite{wingreenAndersonModelOut1994,andersSteadyStateCurrentsNanodevices2008,cohenGreenFunctionsRealtime2014PRB,cohenGreenFunctionsRealTime2014PRL,krivenkoDynamicsKondoVoltage2019}.
Within the lead but near the impurity, $x=2a$, this same phenomena can be observed in the difference between the Kondo and non-Kondo LDOS (top left of bottom panel).
As we move further away from the impurity (remaining panels) the effect of the voltage becomes more complicated as the underlying spectrum becomes more oscillatory.
Nevertheless, increasing the voltage consistently suppresses the overall difference between the Kondo and non-Kondo LDOS at all distances.
This correspondence provides some evidence for the relevance of the lead LDOS for Kondo physics on the impurity.
In general, our analysis provides a new experimentally accessible window into the mechanism by which an applied voltage destroys the Kondo effect \cite{kaminskiSuppressionKondoEffect1999, roschKondoEffectQuantum2001}.
One should however mention Ref.~\onlinecite{erpenbeckResolvingNonequilibriumKondo2020} in this context, where it was shown that at low voltages this behavior may be strongly dependent on the choice of observable.

\section{Conclusions}

We demonstrated two approaches for measuring and quantifying the spatial extent of the Kondo cloud.
Following recent experimental work \cite{borzenetsObservationKondoScreening2020} and preceding theoretical predictions \cite{parkHowDirectlyMeasure2013}, we first considered measuring the Kondo cloud by observing the effect of lead perturbations a distance $L$ away from the QD on the width of the zero-bias density of states, which can be accessed by transport experiments.
This width is used as a proxy for the Kondo temperature.
Consistent with previous results, we find oscillations in the Kondo temperature that decay with $L$.
However, in addition to confirming the general scenario seen in Ref.~\onlinecite{borzenetsObservationKondoScreening2020}, we uncover a more detailed picture of non-monotonic spatial dependence of the Kondo temperature oscillations.
The latter may be observed in future experiments with higher spatial resolution and perhaps more structured leads.
In particular, we show that the even--odd dependence on lead site that generates the oscillations can be suppressed and eventually reversed in sign at large $L$ and $U$.

Having established that we reproduce the key existing experimental observations, we discuss how the Kondo cloud could be observed in a complementary way in the same system via STM experiments.
We identify features in the lead LDOS corresponding to the onset of Kondo physics.
We then show that the width of a peak (dip) in the lead LDOS at the Fermi energy decays with distance from the QD and that the length scale of this decay can be used to define a measurement of the Kondo cloud which appears to display the correct scaling behavior.
We also investigate the effects of an applied bias voltage, finding that the suppression of the Kondo resonance on the impurity by a voltage is accompanied by corresponding changes in the lead LDOS.
These effects should be observable with STM techniques.
Measuring the width of the central LDOS peak would require ${\sim}\thinspace\mu\mathrm{eV}$ spectral resolution, a requirement that can already be experimentally satisfied in some experimental setups \cite{schwenkAchievingMeVTunneling2020}.

\appendix

\section{\label{app:derivation}Calculation of Lead Green's Functions}

In this appendix we derive expressions for the lead Green's functions.
We proceed in three steps.
First, we derive a general form for the Dyson equation.
We then apply this result to derive equations for the noninteracting lead Green's functions which appear in equation \ref{eq:hyb} and \ref{eq:leadG}.
Finally, we apply the same techniques to derive equation \ref{eq:leadG} for the local lead Green's function in the presence of the QD.

Consider a noninteracting Hamiltonian $H = H_0 + V$ where both $H_0$ and $V$ are single-particle operators.
The Green's function for this system is given by
\begin{align}
  G(\omega) = \left(\omega I - H\right)^{-1} = \left(\omega I - H_0 - V\right)^{-1},
\end{align}
where $H$, $H_0$, and $V$ are matrices in the single-particle space.
From this we can obtain
\begin{subequations}
  \label{eq:dyson}
  \begin{align}
    &\left(\omega I - H_0 - V\right) G = I, \\
    \implies &\left(I - \left(\omega I - H_0\right)^{-1} V\right) G = \left(\omega I - H_0\right)^{-1}, \\
    \implies &\left(I - gV\right) G = g, \\
    \implies & G = g + g V G,
  \end{align}
\end{subequations}
where $g(\omega) = \left(\omega I - H_0\right)^{-1}$ is the Green's function for $H_0$.
Note that the equation $G = g + G V g$ can be obtained analogously.

Using these results, we can derive an equation for the noninteracting lead Green's function $g_w(\omega) = \left(\omega I - H_w\right)^{-1}$.
The main difficulty in computing $g_w$ is that the leads are semi-infinite so the matrices are infinite dimensional.
To solve this issue, we let $V$ be the operator describing hopping between site $N$ and $N+1$ of the lead.
$H_0 = H_w - V$ is then partitioned into two disconnected blocks, $A$ and $B$; these consist of the first $N$ sites, and the rest of the semi-infinite chain, respectively.
Applying \ref{eq:dyson} and dropping the lead index $w$, we obtain the equations
\begin{align}
  g_{ij} &= \tilde{g}_{ij} + \tilde{g}_{iN} t g_{N+1,j}, \\
  g_{N+1,j} &= 0 + \tilde{g}_{N+1,N+1} t g_{Nj}, \\
  g_{Nj} &= \tilde{g}_{Nj} + \tilde{g}_{NN} t g_{1N+1,j}.
\end{align}
Here, $\tilde{g} = \left(\omega I - H_0\right)^{-1}$ is the Green's function of the lead without $V$, $t$ is the hopping amplitude between sites $N$ and $N+1$ and we assume $i,j \le N$.
In the second equation, we note that the first term is zero because there are no terms in $H_0$ connecting sites on different sides of the partition.
We also note that the Green's function $\tilde{g}_{N+1,N+1}$ is simply the surface Green's function for a uniform semi-infinite chain, which we denote $\mathcal{G}$.
$\mathcal{G}$ can be computed analytically (see chapter 5 of \cite{cuevasMolecularElectronicsIntroduction2017}).
Combining the second and third equations we obtain
\begin{align}
  &g_{N+1,j} = \mathcal{G} t \tilde{g}_{Nj} + \mathcal{G} t \tilde{g}_{NN} t g_{N+1,j}, \\
  \implies &g_{N+1,j} = \frac{t \tilde{g}_{Nj}}{\mathcal{G}^{-1} - t^2 \tilde{g}_{NN}}.
\end{align}
Combining this with the first equation we obtain
\begin{align}\label{eq:leadg}
  g_{ij} = \tilde{g}_{ij} + t^2 \frac{\tilde{g}_{iN} \tilde{g}_{Nj}}{\mathcal{G}^{-1} - t^2 \tilde{g}_{NN}}.
\end{align}
Now, we note that $H_0$ is block diagonal and can therefore be written as $H_0 = H_0^A \oplus H_0^B$.
We write
\begin{align}
  \tilde{g}
  &= \left(\omega I - H_0\right)^{-1} \\
  &= \left(\omega I - H_0^A \oplus H_0^B\right)^{-1} \\
  &= \left(\omega I - H_0^A\right)^{-1} \oplus \left(\omega I - H_0^B\right)^{-1}.
\end{align}
Since we assumed that $i,j \le N$, we have $\tilde{g}_{ij} = \left(\omega I - H_0^A\right)^{-1}$, which is given entirely in terms of finite dimensional matrices and can be directly computed.
Using this procedure we can compute the noninteracting lead Green's functions $g_w(\omega)$ which appear in equation \ref{eq:hyb} for the coupling density and in equation \ref{eq:leadG} for the lead Green's functions in the presence of the QD.

The derivation of equation \ref{eq:leadG} proceeds in much the same way, but is complicated by the fact that the system now contains interactions.
In the presence of interactions, the full Green's function may be written as
\begin{align}
  G(\omega) = \left(\omega I - H_0 - V - \Sigma\right)^{-1},
\end{align}
where $H = H_0 + V$ is the single-particle Hamiltonian and $\Sigma$ is the self-energy generated by the interaction.
Following the same steps as above we again obtain $G = g + g V G$, with the difference that now $g(\omega) = \left(\omega I - H_0 - \Sigma\right)^{-1}$ contains the self-energy.
This amounts to the observation that one can write the Dyson equation relative to an arbitrary part of the single-particle Hamiltonian rather than the interaction self-energy.

To derive equation \ref{eq:leadG}, let $V$ be the operator describing hopping between the impurity (at site 0) and lead $w$.
Without loss of generality we take $w = r$.
Note that $H_0 = H - V$ is then partitioned into two disconnected blocks, $A$ and $B$ consisting of the impurity together with lead $l$, and lead $r$ respectively.
We then obtain the equations
\begin{align}
  G_{ii} &= \hat{g}_{ii} + \hat{g}_{i1} \lambda G_{0i}, \\
  G_{0i} &= 0 + G_{00} \lambda \hat{g}_{1i},
\end{align}
where $G$ is the full Green's function in the presence of the QD, $\hat{g}(\omega) = \left(\omega I - H_0 - \Sigma\right)^{-1}$, and $\lambda$ is the hopping amplitude between the impurity and lead $r$.
Combining these equations we obtain
\begin{align}
  G_{ii} = \hat{g}_{ii} + \lambda^2 \hat{g}_{i1} G_{00} \hat{g}_{1i}.
\end{align}
This reproduces the form of equation \ref{eq:leadG}.
However, $\hat{g}$ still contains the self-energy $\Sigma$, and so some additional simplification is required.
In this model the self-energy is local to the impurity.
Therefore, $\Sigma$ is zero in subspace $B$, which does not contain the impurity.
This means it can be written as $\Sigma = \Sigma^A \oplus \Sigma^B = \Sigma^A \oplus 0$.
Additionally, because of our choice of $V$, the Hamiltonian $H_0$ is partitioned into two disconnected blocks and can be written as $H_0 = H_0^A \oplus H_0^B = H_0^A + H_r$.

Given all this, we can now write:
\begin{align}
  \hat{g}(\omega) &= \left(\omega I - H_0 - \Sigma\right)^{-1} \\
       &= \left(\omega I - H_0^{A} \oplus H_r - \Sigma^A \oplus 0 \right)^{-1} \\
       &= \left(\omega I - H_0^{A} - \Sigma^{A}\right)^{-1} \oplus \left(\omega I - H_r\right)^{-1}.
\end{align}
This implies that for indices $i,j$ in subspace B (lead $r$) we have
\begin{align}
  \hat{g}_{ij}(\omega) = \left[\left(\omega I - H_r\right)^{-1}\right]_{ij} = g_{r,ij}(\omega).
\end{align}
Using this we obtain
\begin{align}
  G_{r,ii}(\omega) = g_{r,ii}(\omega) + \lambda^2 g_{r,i1}(\omega) G_\mathrm{imp}(\omega) g_{r,1i}(\omega),
\end{align}
which finally is equivalent to equation \ref{eq:leadG}.

\section{\label{app:inch}Inchworm Benchmark}
\begin{figure}
  \includegraphics[]{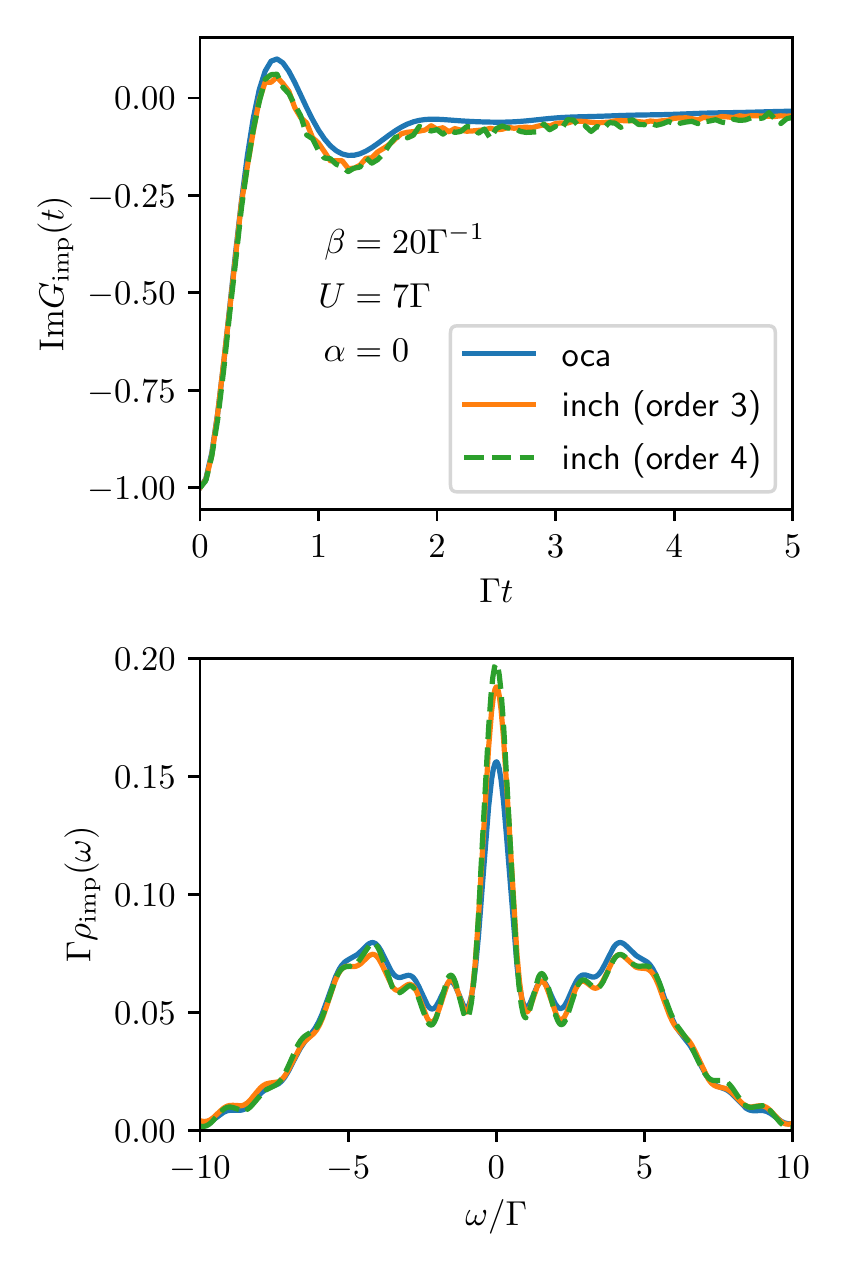}
  \caption{\label{fig:inch} Comparison of the time-dependent retarded QD Green's function (top) and QD DOS (bottom) obtained from OCA (blue) and inchworm QMC (orange/green) at $\beta = 20\Gamma^{-1}$, $U = 7\Gamma$, $\alpha = 0$.
  The inchworm QMC results are parameterized by the maximum allowed expansion order (see legend).
  }
\end{figure}
Fig.~\ref{fig:inch} shows a comparison between results obtained using the one crossing approximation (OCA) and inchworm QMC truncated at order 3 and 4.
Note that the OCA corresponds to inchworm truncated at order 2, and the exact solution is recovered as the truncation order is taken to infinity \cite{cohenTamingDynamicalSign2015, antipovCurrentsGreenFunctions2017}.
The comparison is done at the smallest interaction strength $U = 7\Gamma$ used above, where the OCA is expected to be least accurate.
The inchworm QMC results are converged at orders 3 and 4.
The parameters $t_\mathrm{max} = 5\Gamma^{-1}$ and $\beta = 20\Gamma^{-1}$.
Both the time and inverse temperature are smaller than those used in main text, and we choose them in order to make the inchworm calculations computationally feasible while still making a meaningful comparison.
The unit of time is given by $\hbar / \Gamma \approx 6.5\thinspace\mathrm{ps}$.
The temperature $T = 1/\beta = 0.05\Gamma$ is below the Kondo temperature estimated from the OCA peak width, $T_K \approx 0.075 \Gamma$.
The top panel shows the imaginary part of the time-dependent, retarded QD Green's function $G_\mathrm{imp}(t)$, which is the direct output (before the Fourier transform) in both methods.
The bottom panel shows the DOS, as obtained via a Fourier transform of $G_\mathrm{imp}(t)$.
Note that the DOS is broadened significantly relative to the results in the main text due to the short maximum simulation time $t_\mathrm{max}$, but nevertheless it is clear that the dynamics provided by OCA are almost quantitatively, and certainly qualitatively, reliable in this regime.

\bibliographystyle{apsrev4-2}
\bibliography{refs}

\end{document}